\documentclass[reprint,aip,pop]{revtex4-1}

\usepackage{amsmath}
\usepackage{amssymb}
\usepackage{graphicx}
\usepackage{color}

\begin{document}

\title{Saturation of the leading spike growth in backward Raman amplifiers}
\author{V.~M.~Malkin}
\affiliation{Department of Astrophysical Sciences, Princeton University, Princeton, NJ USA 08540}
\author{Z.~Toroker} 
\affiliation{Department of Electrical Engineering, Technion Israel Institute of Technology, Haifa 32000, Israel}
% \affiliation{Princeton Plasma Physics Laboratory, Princeton, NJ USA 08543}
\author{N.~J.~Fisch}
\affiliation{Department of Astrophysical Sciences, Princeton University, Princeton, NJ USA 08540}
%\affiliation{Princeton Plasma Physics Laboratory, Princeton, NJ USA 08543}

\date{\today}

\begin{abstract}
Backward Raman amplification of laser pulses in plasmas can produce nearly relativistic unfocused output intensities and multi-exawatt powers in compact devices. 
The largest achievable intensity depends on which of major competitive processes sets this limit. 
It is shown here that the relativistic electron nonlinearity can cause saturation of the leading amplified spike intensity before filamentation instabilities develop. 
A simple analytical model for the saturation, which supports numerical simulations, is suggested. 
The upper limit for the leading output spike unfocused intensity is calculated.
\end{abstract}


\maketitle

\section{Introduction}
The highest laser pulse powers are now achieved through the technique of chirped pulse amplification (CPA)  
\cite{Mourou85,Mourou98,Yakovlev_14}, whereby  short laser pulses are stretched, amplified, and re-contracted.  
The gratings needed  can tolerate laser intensities only up to few terawatt  ($10^{12}\,$W) per cm$^2$, which limits  the powers that can be reached with tabletop apertures to few petawatt ($10^{15}\,$W),  for apertures $30\times 30$ cm$^2$.
This limit might be significantly exceeded through the resonant backward Raman amplification (BRA) in plasma  \cite{Malkin_99_PRL}.   

In the BRA scheme, a short seed laser pulse (of frequency $\omega_b$), down-shifted from a much longer counter-propagating pump pulse (of frequency $\omega_a$) by the plasma frequency $\omega_e$,  absorbs the pump energy through a resonant three-wave interaction.
At an advanced nonlinear stage of the pump backscattering instability, 
the amplified pulse shape approaches the classical ``$\pi$-pulse'' wavetrain. 
The leading spike of the``$\pi$-pulse'' wavetrain is dominant; behind the few first spikes, the pump is nearly completely depleted. 
The leading spike grows and contracts until the electron quiver velocities become mildly relativistic, whereupon the plasma is terminated before deleterious instabilities develop.    
The output laser pulse can then be focused within a vacuum region. 
The unfocused output pulse intensities could reach nearly $10^{18}\,$W/cm$^2$ for 1/4$\,\mu$m  wavelength, or $10^{17}\,$W/cm$^2$ % 100 PW/cm$^2$, 
for 0.8$\,\mu$m wavelength lasers. 
The tabletop $30\times 30$ cm$^2$ apertures would then make feasible $10^{21}\,$W laser pulses for 1/4$\,\mu$m  wavelength.   
Experiments %in gas-jet plasmas 
have now demonstrated the possibility of reaching unfocused intensities in backward Raman amplifiers nearly that large \cite{Ping_04_PRL,Balakin_04_JETPL,Cheng_05_PRL,Ren_08_POP, Kirkwood_07_POP, Pai_08_PRL,Jaro_12_NJP}. 

The major physical processes that may affect the amplification include 
the amplified pulse filamentation and detuning due to the relativistic electron nonlinearity \cite{Malkin_99_PRL,Fraiman_02_POP,Malkin_07_PRL,2012-dispersion,Malkin_14-EPJST,PoP-2014-Lehmann}, parasitic Raman scattering of the pump and amplified pulses by plasma noise \cite{Malkin_99_PRL,Malkin_00_PRL,Malkin_00_POP,Tsidulko_00_PRL,Solodov_04_PRE,Malkin_14-EPJST},  generation of superluminous precursors of the amplified pulse \cite{Tsidulko_02_PRL}, pulse scattering by plasma density inhomogeneities \cite{Solodov_dens}, pulse depletion and plasma heating through inverse bremsstrahlung \cite{Malkin_07_PRE,Malkin_09_PRE,Malkin_10_POP,2011-Balakin}, the resonant Langmuir wave Landau damping \cite{PRL-2005-Hur,Malkin_07_PRE,PoP-2009-Yampolsky,Malkin_10_POP,PoP-2011-Yampolsky,PoP-2012-Strozzi,IEEE-2014-Wu,NatCom-2014-Depierreux}, and other, see for example \cite{Clark_03_POP,Yampolsky_04_PRE,Toroker_POP_12, Toroker_12_PRL}.
Most of these deleterious processes can be mitigated by appropriate preparation of laser pulses and plasmas, choosing parameter ranges and selective detuning of the Raman resonance.
Ultimately, the output intensity limit appears to be imposed primarily by the relativistic electron nonlinearity (REN). 
The major goal of this paper is to determine the maximal achievable intensity of the leading amplified spike in REN affected regimes. 

\section{Basic Equations}
To proceed, it is necessary first to outline the relative role of the transverse and longitudinal effects associated with the REN. 

Inasmuch as the amplified pulse consumes incident pump pulse, a transverse non-uniformity of the pump intensity does not affect the amplified pulse, as long as the longitudinally averaged pump intensity does not contain transverse patterns.  In addition the pump intensity is non-relativistic, so that the pump itself will not be affected by relativistic filamentation instabilities.  The amplified pulse, however, reaches much higher intensities, at which point the REN might cause transverse filamentation. This filamentation must be delayed to extend the pulse amplification.
 
At an advanced nonlinear amplification stage, the amplified pulse intensity transverse profile basically reproduces the transverse profile of the longitudinally averaged pump intensity. The amplified pulse transverse profile has also a logarithmically weak dependence from the transverse profile of the longitudinally integrated small input seed amplitude. Both the longitudinally integrated input pump intensity and seed amplitude can be made uniform enough in the transverse direction, so as to keep the parasitically growing transverse modulations benign through many exponentiations.   This will extend the useful amplification until the leading spike saturation limit.  This saturation limit, like the transverse filamentation limit, is imposed by the REN term.

At the largest transverse instability growth rate [\onlinecite{Malkin_99_PRL}], the transverse uniformity needed to reach the saturation limit without the transverse filamentation appears to be such that about 3 exponentiations could be tolerated. This requirement is mild enough, and might be made even milder through the temporarily randomized transverse phase mixing, for which practical means exist [\onlinecite{1989-JournApplPhys-Skupsky}]. 

Note that the non-relativistic laser pulse pump filamentation, whether ponderomotive or thermal, similar to that considered in many papers on the laser driven inertial fusion (see, for example [\onlinecite{Berger-1993-PhysFluidsB}]) can be kept benign in BRA even for long enough pump pulses. This is because the non-relativistic filamentation develops on the ion time scale, while the broad bandwidth of BRA allows to use pump pulse with much shorter time coherence scale, like in the technique [\onlinecite{1989-JournApplPhys-Skupsky}] and simulations  [\onlinecite{Balakin_03_POP}].

Thus, to assess the largest output intensity, a one-dimensional model may be adequate. 
The one-dimensional equations for the resonant 3-wave interaction, together with the lowest order relativistic nonlinearity and group velocity dispersion terms, can be put in the form \cite{Malkin_07_PRL}:
\begin{eqnarray}\displaystyle
&a_t+ c_a a_z=V_3 fb~,\;\;\;\;  f_t=-V_3 ab^*, \label{1} \\
&b_t-c_b b_z=-V_3 af^* +\imath R|b|^2b  -\imath \kappa b_{tt} \, . \label{2}
\end{eqnarray}
Here $a$, $b$ and $f$ are envelopes of the pump pulse, counter-propagating shorter pumped pulse and resonant Langmuir wave, respectively; subscripts $t$ and $z$ signify time and space derivatives; $c_a$ and $c_b$ are group velocities of the pump and amplified pulses;  $V_3$ is the 3-wave coupling constant (real for appropriately defined wave envelopes), $R$ is the coefficient of nonlinear frequency shift due to the relativistic electron nonlinearity, $\kappa=c_b'/2c_b$ is the group velocity dispersion coefficient ($c_b'$ is the derivative of the amplified pulse group velocity over the frequency).  

This hydrodynamic model is applicable for the pump pulse intensity $I_0$ smaller than that at the threshold of the resonant Langmuir wave breaking $I_{\rm br}$. The motivation for studying specifically such regimes is that for deep wavebreaking regimes the BRA efficiency is lower \cite{Malkin_99_PRL}.

The above equations will be solved for a small Gaussian initial seed and constant initial pump with a sharp front.

\section{Specifying coefficients}
The group velocities $c_a$ and $c_b$ are expressed in  terms of the respective laser frequencies $\omega_a$ and $\omega_b$ as follows:
\begin{equation}\displaystyle
c_a=c\sqrt{1-\frac{\omega_e^2}{\omega_a^2}}, \;\quad 
c_b=c\sqrt{1-\frac{\omega_e^2}{\omega_b^2}}\, ,
\label{e3}
\end{equation}
where $c$ is the speed of light in vacuum,
\begin{equation}\displaystyle
\omega_e=\sqrt{\frac{4\pi n_e e^2} {m_e}}
\label{e4}
\end{equation}
is the electron plasma frequency, $n_e$ is the electron plasma concentration, $m_e$ is the electron rest mass and $-e$ is the electron charge),
so that
\begin{equation}\displaystyle
2\kappa=\frac{c_b'}{c_b} = \frac{\omega_e^2}{\omega_b(\omega_b^2-\omega_e^2)}=
%2\kappa\equiv \frac{c_b'}{c_b} = \frac{\omega_e^2}{\omega_b(\omega_b^2-\omega_e^2)}=
\frac{\omega_e^2\, c^2}{\omega_b^3\, c_b^2}  ~. 
\label{e5}
\end{equation} 
The pump pulse envelope, $a$, is further normalized such that the average square of the electron quiver velocity in the pump laser field, measured in units of $c^2$, is $|a|^2$, so that
\begin{equation}\displaystyle
\overline{v_{ea}^2}=c^2|a|^2.
\label{e6}
\end{equation}
Then, the average square of the electron quiver velocity in the seed laser field and in the Langmuir wave field are given by 
\begin{equation}\displaystyle
\overline{v_{eb}^2}=c^2|b|^2 \frac{\omega_a}{\omega_b} , \;\quad
\overline{v_{ef}^2}=c^2|f|^2 \frac{\omega_a}{\omega_f}.
\label{e7}
\end{equation}
The  3-wave coupling constant can be written as \cite{Kruer1988}
\begin{equation}\displaystyle
V_3=k_f c\sqrt{\frac{\omega_e}{8\omega_b}}\, 
\label{e8}
\end{equation}
where $k_f$ is the wave number of the resonant Langmuir wave
\begin{equation}\displaystyle
k_f=k_a+k_b,\;\; k_ac=\sqrt{\omega_a^2-\omega_e^2},\;\; k_bc=\sqrt{\omega_b^2-\omega_e^2}~.
\label{e9}
\end{equation}
The frequency resonance condition is 
\begin{equation}\displaystyle
\omega_b+\omega_f=\omega_a\, ,
\label{e10}
\end{equation}
where $\omega_f\approx \omega_e$ is the Langmuir wave frequency in a cold plasma. 
The nonlinear frequency shift coefficient $R$ can then be put as \cite{Litvak1969,Max1974,Sun1987}
\begin{equation}\displaystyle
R=\frac{\omega_e^2\omega_a}{ 4\omega_b^2}  .
\label{e11}
\end{equation}

\section{Dimensionless variables}
It is convenient to introduce dimensionless times:  $\bar{t}$,  the short seed pulse arrival time to a given location $z$;  and $\tilde{t}$,   the  time elapsed after this arrival:
\begin{equation}\displaystyle
\bar{t}=V_3a_0 \frac{L-z}{c_b}~, \;\;\;\; \tilde{t}= V_3a_0\left(t-\frac{L-z}{c_b}\right),
\label{e12}
\end{equation}
where $L$ is the plasma length in $z$-direction  and $a_0$ is the input pump amplitude, and to define new wave amplitudes as follows: 
$$a=a_0\check{a}\, , \;   f=-a_0\check{f}\, , \; b=a_0\check{b}\, .$$
In these new variables,  Eqs.~(\ref{1}) and (\ref{2}) take the form
\begin{eqnarray}
&\displaystyle \left(1+\frac{c_a}{c_b}\right)\check{a}_{\tilde{t}}-\frac{c_a}{c_b}\check{a}_{\bar{t}} =-\check{f}\check{b}~,\;\;\;\; \check{f}_{\tilde{t}}= \check{a}\check{b}^*, \label{e13} \\
&\displaystyle \check{b}_{\bar{t}} =\check{a}\check{f}^* -\imath \kappa_1 \check{b}_{\tilde{t}\tilde{t}} +\imath R_1|\check{b}|^2\check{b} ~,\label{e14}\\
&\displaystyle \kappa_1=V_3a_0\kappa~, \;\;\;\; R_1=Ra_0/V_3~.
\label{e15}
\end{eqnarray}

\section{Linear amplification regime}
During the linear amplification stage, and also at the nonlinear stage ahead of the leading amplified spike, where the pump depletion and cubic nonlinearity are negligible, these equations reduce to:
\begin{equation}\displaystyle
\check{f}^*_{\tilde{t}}= \check{b},\;\;\;\; 
\check{b}_{\bar{t}} =\check{f}^* -\imath \kappa_1 \check{b}_{\tilde{t}\tilde{t}}~.\label{e16}
\end{equation}
The solution is
\begin{eqnarray}
&\displaystyle \check{b}(\tilde{t},\bar{t})=\partial\int d\tilde{t}_1\check{b}(\tilde{t}_1,0)Y(\tilde{t}-\tilde{t}_1,\bar{t})/\partial\tilde{t}~,\label{e17}\\
&\displaystyle Y(\tilde{t},\bar{t})= \int_0^{2\pi}\frac{d\phi}{2\pi}\exp\left[2\sqrt{\tilde{t}\bar{t}}\cos\phi-
\imath\kappa_1\frac{\bar{t}^2}{\tilde{t}}\exp(2\imath\phi)\right]~.
\label{e18}
\end{eqnarray}
This linear solution will be used below, complimentary to the Eqs. (\ref{e24}-\ref{e25}) of the section VII, to determine how much behind the original seed pulse is located the soliton approximating the leading amplified spike in the nonlinear dispersionless regime. For a small seed, the leading amplified spike is many linear exponentiations behind the seed, so that an asymptotic formula easily derived from (\ref{e17}-\ref{e18}) can be used. 

\section{Advanced nonlinear regime}
At an advanced amplification stage, when the pulse duration is short compared to the amplification time,  so that 
$$|\check{a}_{\tilde{t}}|\gg|\check{a}_{\bar{t}}|$$
  in  Eq. (\ref{e13}),  Eqs.~(\ref{e13}) and (\ref{e14}) can be reduced to
\begin{eqnarray}
&\displaystyle r=\sqrt{1+\frac{c_a}{c_b}},\;\;\; R_2=\frac{R_1}{r} , \;\;\; b_1= R_2^{1/3}\check{b}, \nonumber\\
&\displaystyle f_1=\frac{\check{f}}{r} , \;\;\; \zeta=\frac{\tilde{t}}{rR_2^{1/3}},\;\;\; \tau= rR_2^{1/3}\bar{t},\label{19}\\
&\displaystyle \check{a}_{\zeta}=-b_1 f_1,\;\;\; f_{1\zeta}=\check{a}b_1^*,\label{20}\\
&\displaystyle b_{1\tau}=\check{a} f_1^*-\imath Q b_{1\zeta\zeta}+
\imath |b_1|^2 b_1,\label{21}\\
&\displaystyle Q=\frac{\kappa_1}{r^2R_1}=\frac{(k_a+k_b)^2c^2\omega_b c_b'}{4\omega_e\omega_a(c_a+c_b)}.\label{22}
\end{eqnarray}
In strongly undercritical plasmas, where the plasma frequency is much smaller than the laser frequency, $q=\omega_e/\omega_b \ll 1$, which case is of the major interest here, the amplified pulse intensity $I$ 
can be expressed in these variables as 
\begin{eqnarray}
&\displaystyle I=\frac{G|b_1|^2\omega_e}{4\lambda_b}\left(\frac{I_0^2}{2I_{\rm br}^2}\right)^{1/3}, %\;\;\; 
\label{23}
\end{eqnarray}
where 
\begin{equation}\displaystyle
G=\frac{m_e^2c^4}{e^2}=0.3 \,{\rm \frac{J}{cm}},\;\;\;\;  \lambda_b=\frac{2\pi}{k_b},
\label{e24}
\end{equation}
and 
$$I_{\rm br}=\displaystyle n_em_ec^3q/16$$ 
is the threshold pump intensity for resonant Langmuir wave breaking.   

\section{Leading spike amplitude for {\it Q} = 0}
Eqs.~(\ref{20}) and  (\ref{21}) will be solved now for small input Gaussian seed pulses of the form $$\displaystyle b_1(\zeta,0)= \frac{b_{1_0}}{\sqrt{D\pi}}
\exp\left[-\frac{(\zeta-\zeta_{0})^2}{D}\right]$$ 
with $b_{1_0}=0.05$, $D=1$ and $\zeta_{0}=10$. 
No auxiliary chirping of the seed pulse is needed here, though it may be useful in less undercritical plasmas \cite{Toroker_12_PRL}.

\begin{figure}[ht]
\includegraphics[width=0.5 \textwidth]{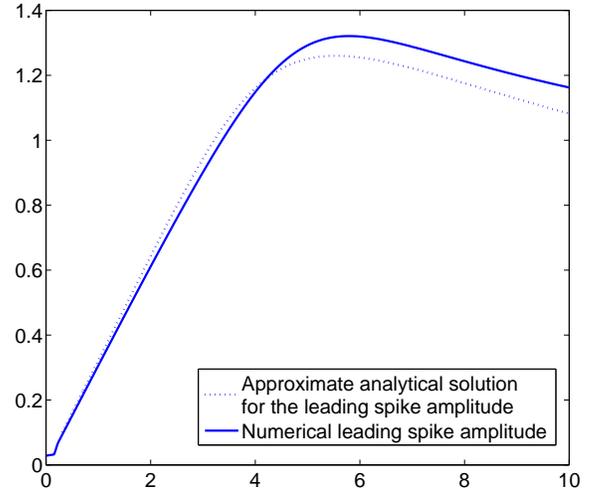}
\caption{The leading spike amplitude $\max_\zeta |b_1|$ as a function of the amplification time $\tau$ for $Q=0$.} 
\label{f2} \end{figure}

First, consider extremely undercritical plasmas where the group velocity dispersion can be neglected, so that the approximation $Q=0$ is good enough. 
The leading spike amplitude $\max_\zeta |b_1|$ as a function of the amplification time $\tau$, calculated numerically, is shown in the Fig.~\ref{f2}. The nearly straight initial part of the curve represents the classical $\pi$-pulse regime.   In the absence of relativistic electron nonlinearity, this straight part would continue until the applicability limit of the time-envelope equation for the Langmuir wave amplitude. 

The figure also shows an approximate analytical solution. It is obtained as follows.
It can be seen that, for a small seed pulse $|b_{1_0} |\ll 1$, the distance from the original seed to the leading amplified spike is much larger than the spike width. 
Therefore the spike top hardly changes while crossing its own width and can be approximated by a quasi-stationary moving soliton.  
This approximation is  similar to approximating the leading spike of  $\pi$-pulse wavetrain by a quasi-stationary moving $2\pi$-pulse \cite{Malkin_99_PRL}. 
More specifically, in the REN  regime, the growing and contracting nonlinear solution, having the linear asymptotics (\ref{e17}) ahead of the leading amplified spike, can be approximated near the top of this spike as follows:
\begin{eqnarray}
&\displaystyle \xi=2\sqrt{\tau(\zeta-\zeta_{0})}, \;\;\;\; \check{a}=A(\xi-\xi_M),\nonumber \\
&f_1=F(\xi-\xi_M),\;\;\;\; b_1=2\tau B (\xi-\xi_M)/\xi_M, \label{e24}\\
&A'=-F B,\;\;\;\; F'=A B^*, \nonumber\\
&B'=A F^* + \imath s |B|^2B, \;\;\;\; 
 s= \left(2\tau/\xi_M\right)^3 , \label{e25}
\end{eqnarray}
where $\xi_M\gg 1$ is the location of the maximum of the classical $2\pi$-pulse,
\begin{equation}\displaystyle
B=\frac{1}{\cosh(\xi-\xi_M)}=F, \;\; A=-\tanh(\xi-\xi_M),
\label{e26}
\end{equation}
solving Eq.~(\ref{e25}) for $s=Q=0$. 
The maximum  of the $2\pi$-pulse occurs at 
$\xi_M=\log (4\sqrt{2\pi\xi_M}/b_{1_0})$.
 For $b_{1_0}=0.05$,  it gives $\xi_M=6.2$.
 Solutions of Eq.~(\ref{e25}), depending on the parameter $s$ (or $\tau$), generalize  the  $2\pi$-pulse (\ref{e26}) (corresponding to $s=0$).
Like the $2\pi$-pulse, these solutions satisfy conditions $|F|=|B|$ and $|A|=\sqrt{1-|B|^2}$.

The dotted line in Fig. \ref{f2} shows the $\tau$-dependence of the leading spike amplitude $|b_1|=s^{1/3}|B|$,   calculated using Eqs.~(\ref{e24}) and (\ref{e25}) and  maximized over $\xi$.    
The  top amplitude of the analytically approximated leading spike, $\max_s (s^{1/3}\max_\xi{|B|})=1.26$, is reached at %$s^{1/3}=1.78$ (
$s=5.64$  (where $\max_\xi{|B|}=1/\sqrt{2}$). As seen, even such a crude zero-order analytic approximation agrees reasonably well with the numerical solution shown by the solid line in Fig. \ref{f2}.
The initial nearly linear part of the curve  in   Fig.~\ref{f2} corresponds to the classical $\pi$-pulse regime. In the REN regime, the leading spike growth saturates. 

\section{Group velocity dispersion effect}

\begin{figure}[ht]
%\vskip-0.5cm
\includegraphics[width=0.5 \textwidth]{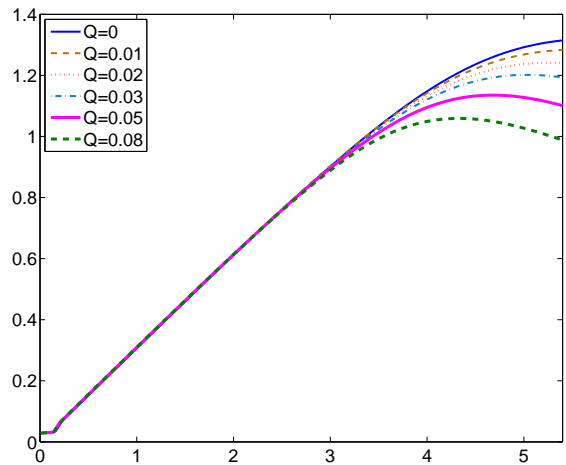}
% replace with shorter name of figure 
%\vskip-0.45cm%0.29in
%\includegraphics[width=0.5 \textwidth]{nln2pipls-time-dpndnc-k01-03-05--num-adv.eps}  
%\includegraphics[width=0.28 \textwidth]{nln2pipls-Fig4b.eps}
%\vskip-0.5cm
\caption{The maximal amplitude of the amplified  pulse $\max_\zeta |b_1|$ as a function of the amplification time $\tau$ for several $Q$.}
\label{f4} 
\end{figure}

For less undercritical plasma, group velocity dispersion begins to play a role. In contrast to the $\pi$-pulse regime, which is not much affected by the group velocity dispersion as long as the plasma remains strongly undercritical, the REN regime can be substantially modified  by the dispersion even in strongly undercritical plasmas. 
This is because the amplification of the leading spike much slows down in the REN regime, as the growth approaches to the saturation, which gives more time for the dispersion to manifest.
That  even  rather small group velocity dispersion  is important is illustrated in Fig.~\ref{f4}. It shows  the maximal amplified pulse amplitude $\max_\zeta |b_1|$ as a function of the amplification time $\tau$.  The dispersion is characterized therein by the parameter {\it Q}, which depends only on the ratio of the plasma to laser frequency  ($q \equiv\omega_e/\omega_b$),   see Eq.~(\ref{22}).  In strongly undercritical plasmas, where $q\ll 1$, one has $Q=q/2$.  In nearly critical plasmas, where $q\rightarrow 1$, one has $Q=0.5/\sqrt{1-q^2}\gg 1$.  Fig.~\ref{f4} illustrates the dispersion effect at small $Q\approx q/2$. The figure shows both the $\pi$-pulse regime and the REN regimes up to the times when saturation occurs, which are of the most interest here. 
The $\pi$-pulse regime corresponds to the joint straight part of the curves.  
The absence of a {\it Q}-dependence there indicates the negligibility of the group velocity dispersion effect.  
However, in the REN regime, the {\it Q}-dependence becomes noticeable. 
The maximal pulse amplitude decreases there, when  $Q$ increases. This is because the group velocity dispersion tends to stretch the pulses.  
It also tends to reduce the saturation time.

Note that the largest growth factor for parasitic transverse modulations the amplification time $\tau$ can be evaluated \cite{Malkin_99_PRL} as $\int d\tau \max_\zeta |b_1|^2$.

Note also that the plasma concentration practically allowed here is bounded from below by the requirement that the Langmuir wave phase velocity, $c_f=\omega_f/k_f\approx Q\,c$, be larger than the electron thermal velocity. The energy of an electron moving with the Langmuir wave phase velocity constitutes $Q^2/2$ fraction of the electron rest energy. For $Q=0.01$, the energy of an electron moving with the Langmuir wave phase velocity is 25 eV. For the electron plasma temperature 10 eV, both the collisional and Landau damping of such a Langmuir wave are small enough for not affecting the BRA regime considered here. Thus, even such small as 0.01 values of the parameter $Q$ are quite practical.

\section{Summary}
It is shown that the relativistic electron nonlinearity saturates  growth of the leading spike in backward Raman amplifiers before the longitudinal filamentation instability of this spike develops.
The time dependence of the maximal amplitude of the leading amplified spike $\max_\zeta |b_1|$ is calculated  for various small plasma-to-laser frequency ratios, as shown in Fig.~\ref{f4}. 
Together with the formula (\ref{23}), this determines the largest unfocused intensity of leading amplified spike achievable in backward Raman amplifiers.

For example, for  $\lambda_b=1/4\,\mu$m and $I_0=I_{\rm br}/2$, and $Q=0.025$ (corresponding to $\omega_e/\omega_b\approx 0.05$), the largest intensity is $8\times 10^{17}\,\rm W/cm^2$.
In this example, the plasma concentration is $n_e=4.5\times 10^{19}\,\rm cm^{-3}$,  the input pump intensity is $I_0=1.7\times 10^{14}\,\rm W/cm^2$ and the pump duration is 0.2 ns.
The ability to compress 0.2 ns laser pulses may allow direct BRA of currently available powerful 1/4 micron wavelength laser pulses to ultrahigh powers.
The regimes found here can further enhance multi-step BRA schemes \cite{Fisch_03_POP, Malkin_05_POP}, as well as possible combinations of such schemes with other currently considered methods of producing ultra-high laser intensities, like \cite{Shvets_98_PRL,PRL-2010-Lancia,UFN-2011-Korzhimanov,RevModPhys-2012-Piazza, Mourou_2012,UFN-2013-Bulanov,PRL-2013-Weber,PRL-2014-Tamburini}.

Note that the largest output intensity here is about 10 times larger than intensities reached in the particle-in-cell (PIC) simulations \cite{Trines_11_PRL}. 
The REN regime was not apparently reached in these simulations. 
This is likely due to the premature backscattering of the pump by numerical noise of the PIC code, which would not be physical noise. Such a parasitic backscattering, whether due to numerical noise or due to physical noise, might have been suppressed (along with few other major
parasitic processes), say, by applying selective detuning techniques
\cite{Malkin_00_PRL,Malkin_00_POP,Tsidulko_00_PRL,Solodov_04_PRE,Malkin_14-EPJST}.
 However, these detuning techniques were not employed in
\cite{Trines_11_PRL}.

All the above addresses only the backward Raman amplification regimes without strong Langmuir wave breaking.
We do not examine here a possibility of an extension of these REN regimes to much higher pump intensities $I_0\gg I_{\rm br}$, where surprisingly large output intensities, like $4\times 10^{17}\,\rm W/cm^2$ for $I_0\approx 30 \, I_{\rm br}$, were reported numerically \cite{Trines_11}.
The underlying mechanisms of surprisingly large BRA efficiency reported in \cite{Trines_11} for these regimes have not yet been identified, and thus require both further investigation and better understanding.

\section{Acknowledgments}
This work was supported by DTRA 
HDTRA1-11-1-0037, by NSF PHY-1202162, and by
the NNSA SSAA Program under Grant No~ DE274-FG52-08NA28553.
%\newpage
%\bibliography{bibBRA2014}
%\bibliographystyle{phaip}
%\bibliographystyle{osa}
%\end{document}

\providecommand{\noopsort}[1]{}\providecommand{\singleletter}[1]{#1}%

\end{document}